\documentclass[prb,showpacs,twocolumn,floatfix, groupaddress,superscriptaddress,footinbib]{revtex4-1}
\usepackage{times,amsmath,amsfonts,amssymb,latexsym}
\usepackage{graphicx,epsf}
\usepackage{subfigure}
\usepackage[ruled,vlined]{algorithm2e}
\usepackage{color}
\newcommand{\braket}[2]{\left \langle #1 | #2 \right\rangle}

\newcommand{\be}{\begin{equation}}
\newcommand{\ee}{\end{equation}}
\newcommand{\ba}{\begin{eqnarray}}
\newcommand{\ea}{\end{eqnarray}}

\newcommand{\ignore}[1]{}

\newcommand{\ket}[1]{\left | {#1} \right \rangle }

\newcommand{\bra}[1]{\left \langle {#1} \right | }

\def\CC{{\rm\kern.24em \vrule width.04em height1.46ex depth-.07ex
    \kern-.30em C}}
\def\P{{\rm I\kern-.25em P}}
\def\RR{{\rm
         \vrule width.04em height1.58ex depth-.0ex
         \kern-.04em R}}

\def\bbbc{{\mathchoice {\setbox0=\hbox{$\displaystyle\rm C$}\hbox{\hbox
to0pt{\kern0.4\wd0\vrule height0.9\ht0\hss}\box0}}
{\setbox0=\hbox{$\textstyle\rm C$}\hbox{\hbox
to0pt{\kern0.4\wd0\vrule height0.9\ht0\hss}\box0}}
{\setbox0=\hbox{$\scriptstyle\rm C$}\hbox{\hbox
to0pt{\kern0.4\wd0\vrule height0.9\ht0\hss}\box0}}
{\setbox0=\hbox{$\scriptscriptstyle\rm C$}\hbox{\hbox
to0pt{\kern0.4\wd0\vrule height0.9\ht0\hss}\box0}}}}
\def\bbbz{{\mathchoice {\hbox{$\sf\textstyle Z\kern-0.4em Z$}}
{\hbox{$\sf\textstyle Z\kern-0.4em Z$}}
{\hbox{$\sf\scriptstyle Z\kern-0.3em Z$}}
{\hbox{$\sf\scriptscriptstyle Z\kern-0.2em Z$}}}}

\usepackage{hyperref}

\begin{document}


\title{Quantum Coherence in Ergodic and Many-Body Localized Systems}
\author{Sayandip Dhara}
\affiliation{Department of Physics, University of Central Florida, Orlando, Florida 32816, USA}
\author{Alioscia Hamma}
\affiliation{Physics department, University of Massachusetts, Boston, Massachusetts 02125, USA}
\author{Eduardo R. Mucciolo}
\affiliation{Department of Physics, University of Central Florida, Orlando, Florida 32816, USA}

\date{\today}

\begin{abstract}
Quantum coherence quantifies the amount of superposition a quantum
state can have in a given basis. Since there is a difference in the
structure of eigenstates of the ergodic and many-body localized
systems, we expect them also to differ in terms of their coherences in
a given basis. Here, we numerically calculate different measures of
quantum coherence in the excited eigenstates of an interacting
disordered Hamiltonian as a function of the disorder. We show that
quantum coherence can be used as an order parameter to detect the
well-studied ergodic to many-body-localized phase transition. We also
perform quantum quench studies to distinguish the behavior of
coherence in thermalized and localized phases. We then present a
protocol to calculate measurement-based localizable coherence to
investigate the thermal and many-body localized phases. The protocol
allows one to investigate quantum correlations experimentally in a
non-destructive way, in contrast to measures that require tracing out
a subsystem, which always destroys coherence and correlation.
\end{abstract}

\maketitle


\section{Introduction}

Advances in experimental realizations of closed quantum many-body
systems such as ultra-cold atoms, trapped ions, or superconducting
qubits undergoing unitary evolution over long time scales
\cite{exp1,exp2,exp3,exp4,exp5} have lead to the study of quantum
dynamical phenomena like dynamical quantum phase
transitions,\cite{dyn1,dyn2} discrete time crystals,\cite{timecr} and
many-body localization (MBL).\cite{mbl}
One of the main focuses of these studies is to examine the way
isolated systems reach thermal equilibrium. Deutsch \cite{Deutsch} and
Srednicki \cite{Sred} discussed the process of thermalization and the
eigenstate thermalization hypothesis (ETH) \cite{Sred} was put forward
as a strong criterion for thermalization to occur in closed quantum
many-body systems. MBL \cite{BASKO,poyla,Huse,NandRe,Pal,RG,Ponte} has
emerged as an extension of the much-studied Anderson
localization,\cite{Anderson} applicable in the case of closed
interacting systems. MBL systems fail to thermalize due to the
presence of local integrals of motion \cite{LIOM,Huseph,Serbyn} and
hence the MBL eigenstates violate the ETH hypothesis, according to
which all the eigenstates of a thermalizing system have to be locally
thermal. ETH also postulates that the matrix elements of any local
observable $O$, between two eigenstates $i,j$ of the Hamiltonian can
be expressed as $\bra{i}{O}\ket{j} = O(\bar{E})\delta_{ij} +
\exp(-S(\bar{E})/2)f_{O}(\bar{E},\omega)R_{ij}$, where $\bar{E}\equiv
(E_{i}+E_{j})/2, \omega = E_{j}-E_{i},$ and $S(E)$ is the
thermodynamic entropy at energy $E$. It is also important to note that
both $O(\bar{E})$ and $f_{O}(\bar{E},\omega)$ are smooth functions of
their arguments and $R_{ij}$ is a random real or complex variable with
zero mean and unit variance.

The effort of keeping a quantum system decoupled from the environment
and thus undergoing unitary dynamics is done with the goal of
preserving coherence in the many-body wave-function. Coherence
quantifies the amount of superposition of a particular state in any
fixed basis sets. A rigorous framework for quantum coherence as a
resource has been developed
recently.\cite{cohresource,ecoh1,ecoh2,ecoh3} The study of quantum
coherence in closed quantum systems is relevant because quantum
coherence is exactly what is responsible for quantum fluctuations and
correlations.
In a many-body quantum system, local degrees of freedom are described
by a tensor product structure (TPS). Coherent superposition of basis
states in a TPS results in quantum entanglement and this is why, in
recent years, entanglement has been widely studied as a diagnostic
tool for quantum phase transitions in many-body systems
\cite{qmb_review} or as a probe to exotic quantum orders like
topological
order.\cite{toporder1,toporder2,toporder3,toporder4,toporder5, kitaev,
  levin} In the context of the ETH-MBL phase transition in spin
chains, entanglement has been used as a useful marker of the
transition.\cite{edmbl,lsmbl,mucc1,umblt} In quantum many-body
dynamics, the nature of the growth of entanglement entropy has been
considered as an important tool for characterizing different dynamical
phases. It has been shown that MBL offers slow logarithmic growth
while ETH has a linear growth of entanglement entropy.\cite{Ent1,Ent2}
 
In this paper, we study the role that quantum coherence plays in the
MBL-ETH transition. As coherence is a function of the wave function,
one should expect that some of its moments should be able to capture
any kind of transition. We first show that coherence (in the
computational basis) in a high energy eigenstate and its variance due
to sample-to-sample fluctuations do indeed signal the MBL-ETH
transition. Second, we look at the coherence/decoherence power of
dynamics generated by MBL and ETH Hamiltonians. We find that ergodic
dynamics induced by ETH has more coherence/decoherence power in a
basis that is incompatible with that of the energy, while the dynamics
induced by MBL has a low coherence/decoherence power, or, in other
words, retains memory of the initial conditions.
 
However, quantum coherence does not contain any information about the
TPS and is, by itself, useless to discriminate the localized versus
unlocalized structure of quantum states. To this end, we exploit the
notion of localizable coherence that has recently been put forward in
Ref. \onlinecite{loc_coh}. Localizing coherence to two blocks of
spins, we can then compute the coherence in these two blocks as a
function of their distance $d(A,B)$, as a coherence-connected
correlation function $C_d$. We show that while this quantity does
depend on $d$ within the dynamics induced by a MBL Hamiltonian, the
ergodic dynamics induced by the ETH Hamiltonian is insensitive to the
distance between the two blocks. We finally note that due to the
projective nature of coherence measures, they are more suitable to
experimental investigation than entanglement entropy, making our
results amenable to testing beyond numerical computations.

\section{Measures of quantum coherence}

Quantum coherence is a notion relative to a specific basis. A
(Hermitian) operator is called incoherent if it is diagonal in a
particular basis $B = \{\ket{i}\}$. We call $I_B$ the set of
incoherent states in $B$. As an example, the Gibbs state is incoherent
in the energy eigenbasis $E$ since its completely diagonal in
it. Every completely dephased operator in $B$ is also incoherent in
that basis. The set $I_B$ is given by just any probability
distribution over $\pi_i$, where $\pi_{i} = \ket{i}\bra{i}$ are the
projectors in the basis $B$. Thus we can say that any completely
dephased operator $X \in I_B$ can be expressed as $X = \sum_i p_i
\pi_i$. Therefore, a coherence measure for a state $\rho$ is the
quantity
\begin{equation}
C_{B,l_p}(\rho):= \|\rho -\mathcal D_B (\rho )\|_{l_p},
\end{equation}
where $\mathcal D_B (\rho) = \sum_{i} \pi_{i} \rho \pi_{i} $ is the
completely dephased state and the measure is based on the $l_{p}$
norm. According to this definition, a state $\rho$ has zero coherence,
$C_B(\rho)=0$, if and only if $\rho\in I_B$. We use two different
matrix norms as measure for coherence.\cite{cohresource,ecoh3} Using
the $l_1$ norm, coherence is expressed as the sum of all the
off-diagonal elements of the quantum state, that is, $C_{B,l_1}(\rho)
= \sum_{i\ne j} |\langle i|\rho | j\rangle |$. Similarly, using the
$l_2$ norm measure we obtain $C_{B,l_2}(\rho) = \sum_{i\ne j}|\langle
i|\rho | j\rangle |^2$.

Another way of measuring coherence in a basis $B$, which we also
employ, is through the Kullback-Leibler divergence from the completely
dephased state,
\begin{equation}
  \label{eq:KL}
  C_B^{\rm KL} (\rho): = S(\mathcal D_B(\rho)) -S(\rho),
\end{equation}
where $S(\rho)$ indicates the entropy of the state $\rho$.

%
%

\section{Quantum Coherence in disordered spin chain}

In order to study the role of coherence in the ETH-MBL transition, we
consider the disordered Heisenberg $1/2$-spin chain \cite{prosenmbl}
described by the Hamiltonian
\begin{equation}
  \label{eq:H}
  H = \sum_{i=1}^{N} J(S_i^x S_{i+1}^x + S_i^y S_{i+1}^y + S_i^z
  S_{i+1}^z) + h_i S_i^z + h_{x} S_i^x
\end{equation}
with periodic boundary conditions. We set $J = 1$ in the
numerical computation. The static random fields $h_{i}$ are chosen
from a uniform distribution in $[-W,W]$. A transverse constant field
$h_{x} = 0.1$ is introduced to break the total $S_{z}$ conservation so
that no sector with conserved quantities that break ergodicity
explicitly exist.\cite{floqNand,mucc1}

\subsection{ETH-MBL transition point from level statistics}

\begin{figure}
	\centering
	\includegraphics[width = 3.2in]{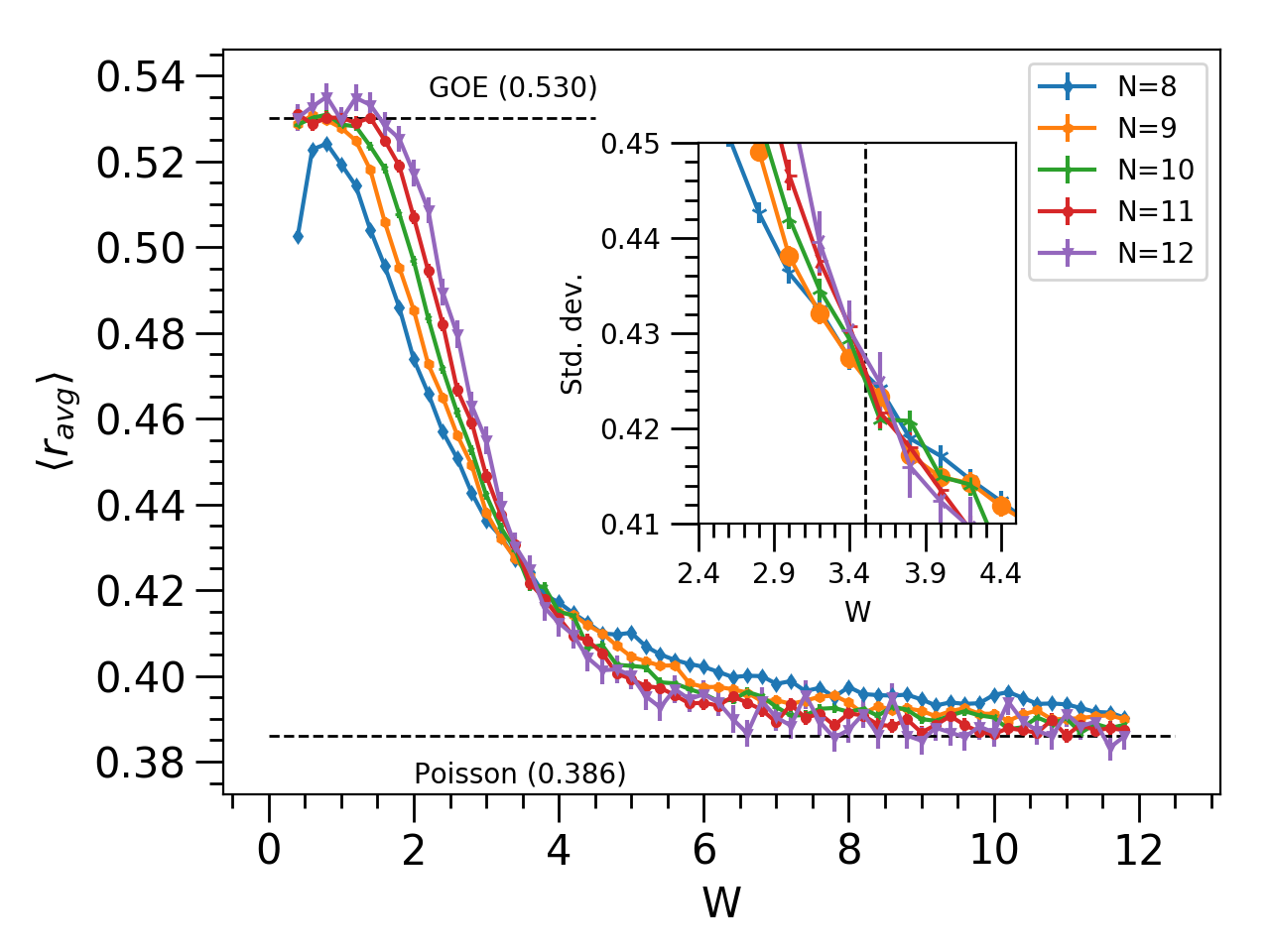}
	\caption{Average level spacing ratio $\langle r_{\rm
            avg}\rangle$ versus the disorder strength $W$ for
          different system sizes ($N$ identifies the number of spins
          in the chain). We obtain $\langle r_{\rm avg}\rangle$ by
          first averaging over $10$ eigenstates near the middle of the
          spectrum for each disorder realization and then averaging
          over different realizations. We employed $8000$ disorder
          samples for $N = 8,9$, $4000$ for $N = 10,11$, and $1000$
          for $N=12$.}
	\label{fig:1}
\end{figure}

The model Hamiltonian in Eq. (\ref{eq:H}) without the small transverse
field ($h_{x}$) is known to undergo an ergodic to MBL transition for
strong disorder.\cite{Pal,lsmbl,edmbl} Here, in order to locate
transition point between the eigenstate thermalization hypothesis
(ETH) to many-body localization (MBL) regimes, we diagonalize the full
Hamiltonian in Eq. (3) and calculate the energy level spacing
$\delta_{\alpha}^{n} = |E_{\alpha}^{n} - E_{\alpha}^{n+1}|$, where
$E_{\alpha}^{n}$ is the energy of the $n$-th eigenstate in the
$\alpha$-th disorder sample. The ratio of the adjacent gaps or level
spacings $r_{\alpha}^{n} = \mbox{min}
\{\delta_{\alpha}^{n},\delta_{\alpha}^{n+1}\} / \mbox{max}
\{\delta_{\alpha}^{n},\delta_{\alpha}^{n+1}\}$ is averaged over the
samples to yield $\langle r_{\rm avg} \rangle$. In random matrix
theory, when the statistical distribution of level spacing follows the
the predictions of the Gaussian Orthogonal Ensemble (GOE) $\langle
r_{\rm avg} \rangle$ converges to $r_{\rm GOE} \approx 0.53$ for $N
\rightarrow \infty$. We find that, deep in the ergodic phase, the
average ratio $\langle r_{\rm avg} \rangle$ does approach the GOE
(Gaussian Orthogonal Ensemble) value (see Fig. \ref{fig:1}). On the
other hand, deep in the localized phase, it reaches the value derived
from a Poisson distribution of level spacings, and $\langle r_{\rm
  avg} \rangle$ converges to $r_{\rm Poisson} \approx
0.39$. Finite-size scaling gives an estimate of the critical value of
disorder to drive the transition from ETH to MBL at $W \approx 3.5$.

\subsection{ETH-MBL transition point from quantum coherence}

\begin{figure}
  \centering
  \includegraphics[width = 3in]{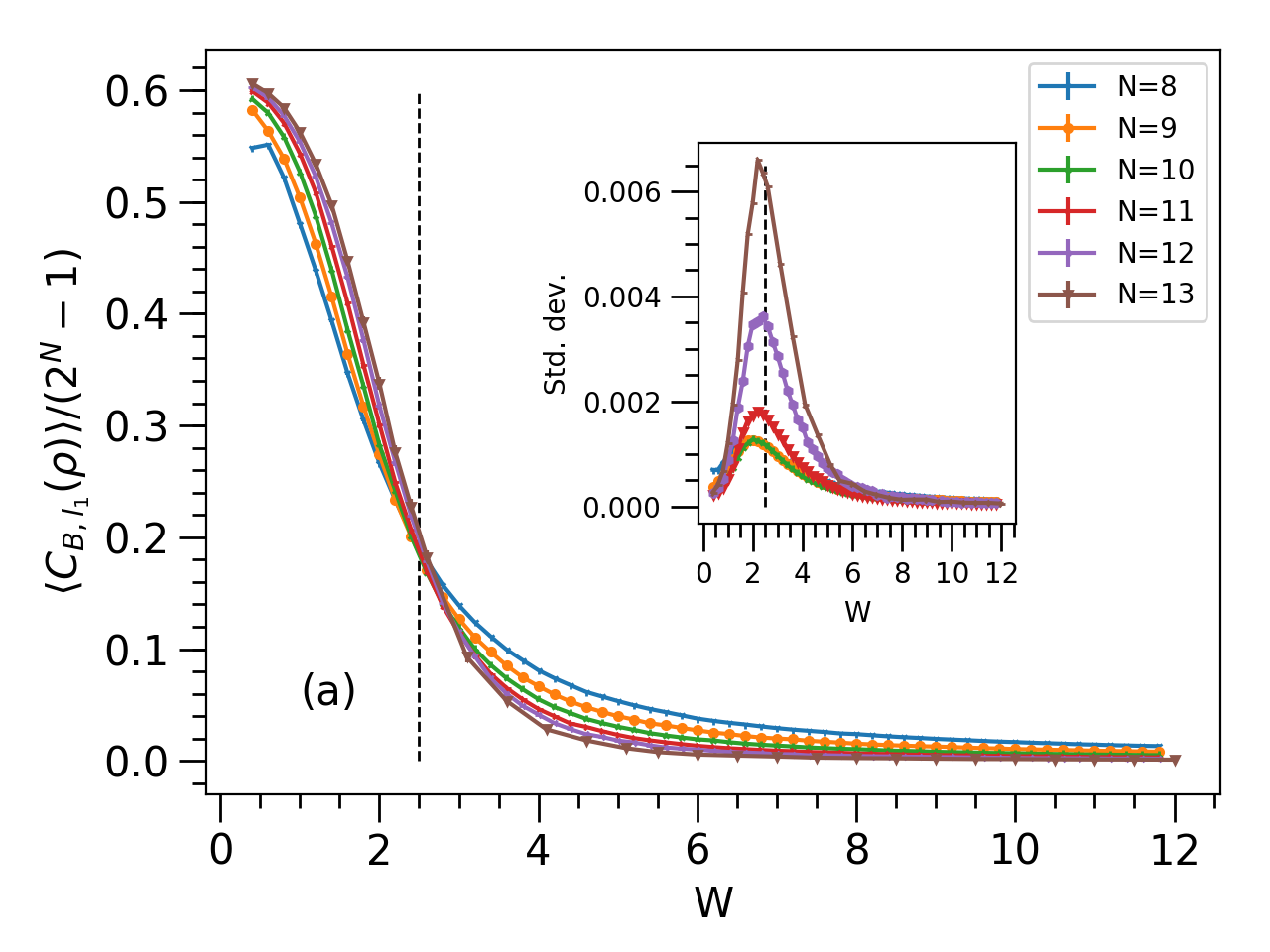}
  \includegraphics[width = 3in]{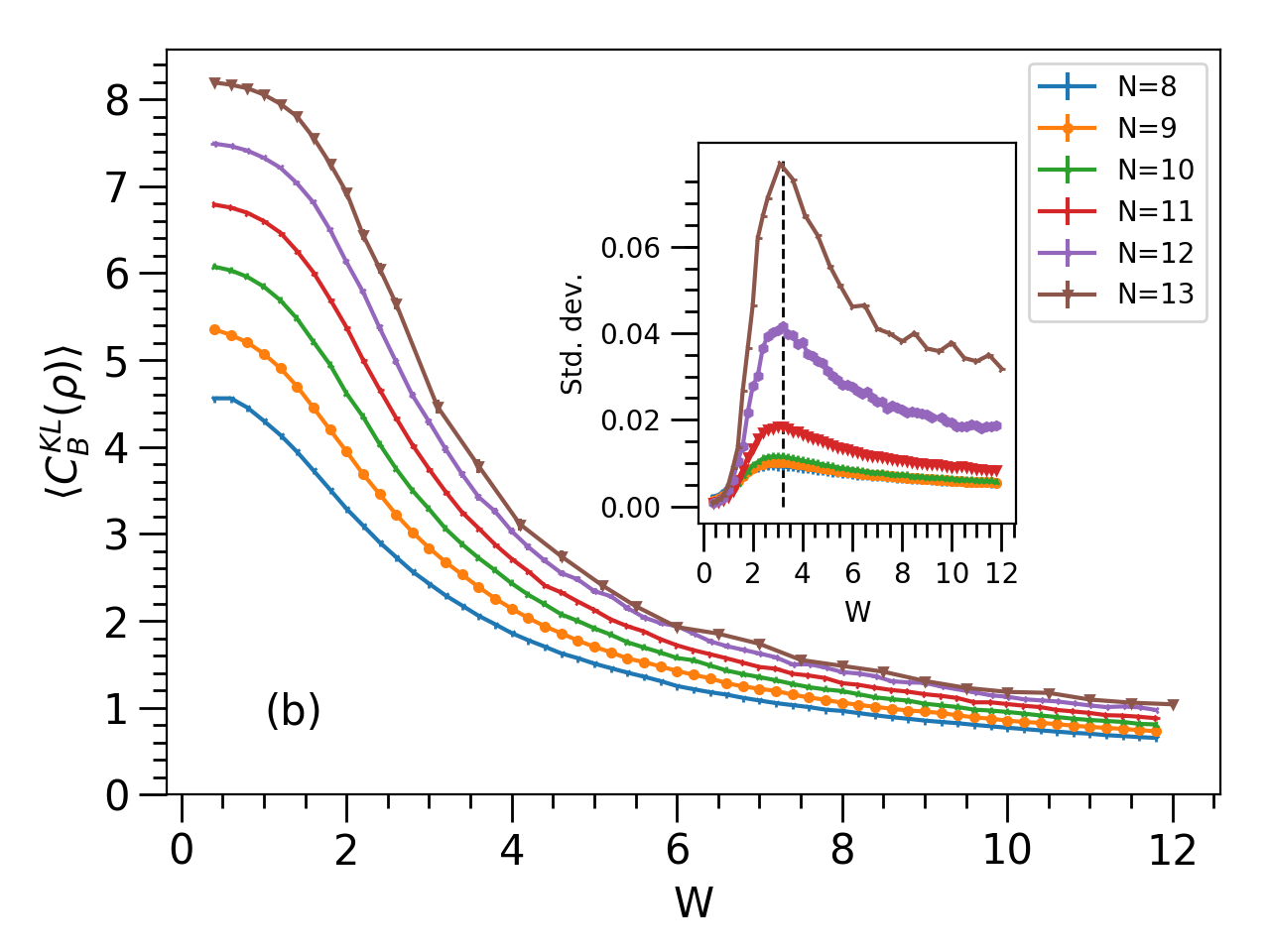}
  \caption{Average normalized coherences: (a) $\langle C_{B,l_1}
    (\rho) \rangle$ and (b) $\langle C_B^{\rm KL} (\rho) \rangle$ of
    an excited state as a function of disorder for different chain
    sizes. We use the eigenstate exactly at the middle of spectrum for
    each case. The data are averaged over $8000$ disorder samples for
    $N = 8, 9$, $4000$ for $N = 10, 11$, and $1000$ for $N=12,
    13$. The inset shows standard deviation of the normalized
    coherence as a function of disorder for the specific eigenstates
    mentioned earlier.}
  \label{fig:2}
\end{figure}

We now show that one can also extract information about this
transition from measures of coherence. Recently,\cite{stzan} it was
shown that the escape probability and dynamical conductivity are
connected by measures of coherence that can effectively probe the
localization transition. Since the ETH and MBL phases are
characterized by the different structures of the high-energy
eigenstates, we start by evaluating the coherence present in a
eigenstate in the middle of the spectrum. As a basis, we choose the
computational ($z$) basis for the tensor product of the spins as the
preferred basis in which one can observe quantum fluctuations. Here we
calculate coherence using $l_1$. The disorder-averaged normalized
coherence $\langle \mbox{Coh} \rangle = \langle C_{B,l_1}(\rho)
\rangle / (2^N-1)$ for different system sizes feature a crossing at a
disorder value around $W = 2.5$, see Fig. \ref{fig:2}a. The standard
deviation of the normalized coherence due to sample-to-sample
variations also shows critical behavior around $W=2.5$, see inset in
Fig. \ref{fig:2}a.


Next, we calculate the average Kullback-Leibler divergence between the
completely dephased state and a high-energy eigenstate, see
Eq. (\ref{eq:KL}). In this case, $\langle C_{B}^{\rm KL}(\rho)
\rangle$ does not reveal any crossing point for different system
sizes, see Fig. \ref{fig:2}b. However, similarly to $\langle
C_{B,l_1}(\rho) \rangle$ in the inset of Fig. \ref{fig:2}a, the
standard deviation does show a well-defined peak, but in this case the
peak is centered at $W =3.2$ for the system sizes we investigated.
Although a peak is much easier to follow and employ for finite-size
scaling analyzes than a line crossing (Fig. \ref{fig:1}), larger
systems would nevertheless be for an accurate estimate of the
transition point location.

\subsection{Coherence after a quantum quench}

\begin{figure}
	\centering
	\includegraphics[width = 3in]{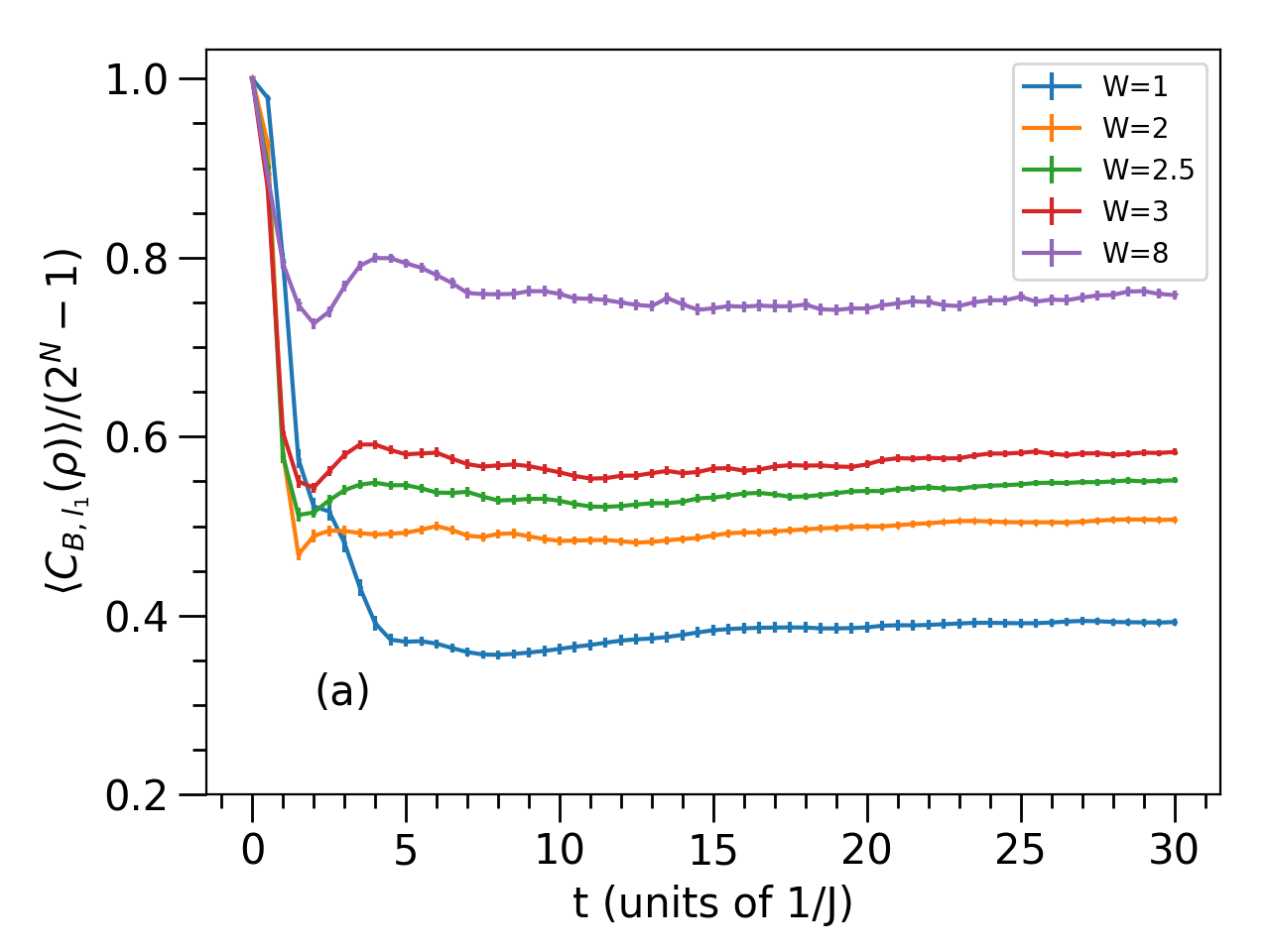}
	\includegraphics[width = 3in]{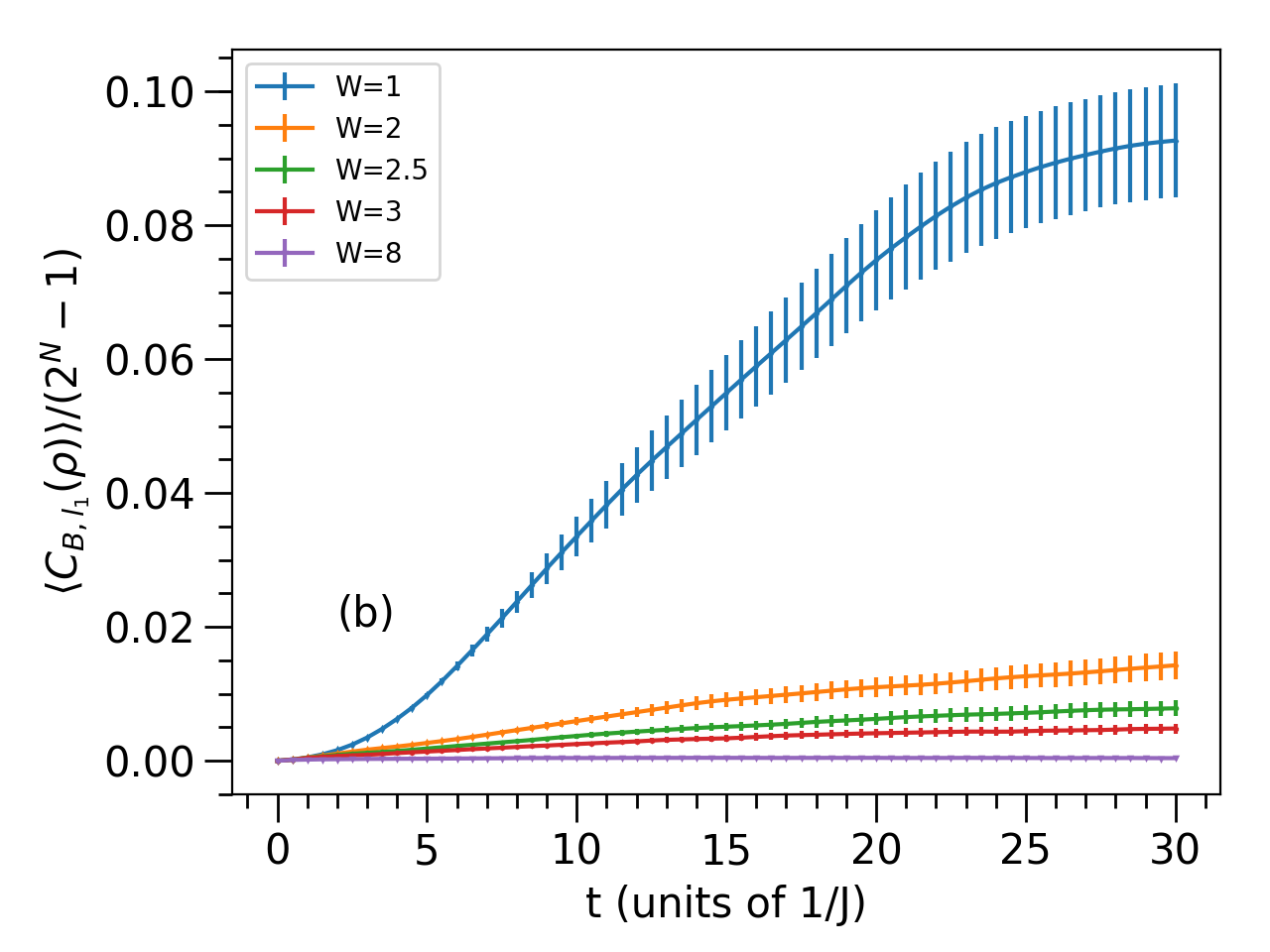}
	\includegraphics[width = 3in]{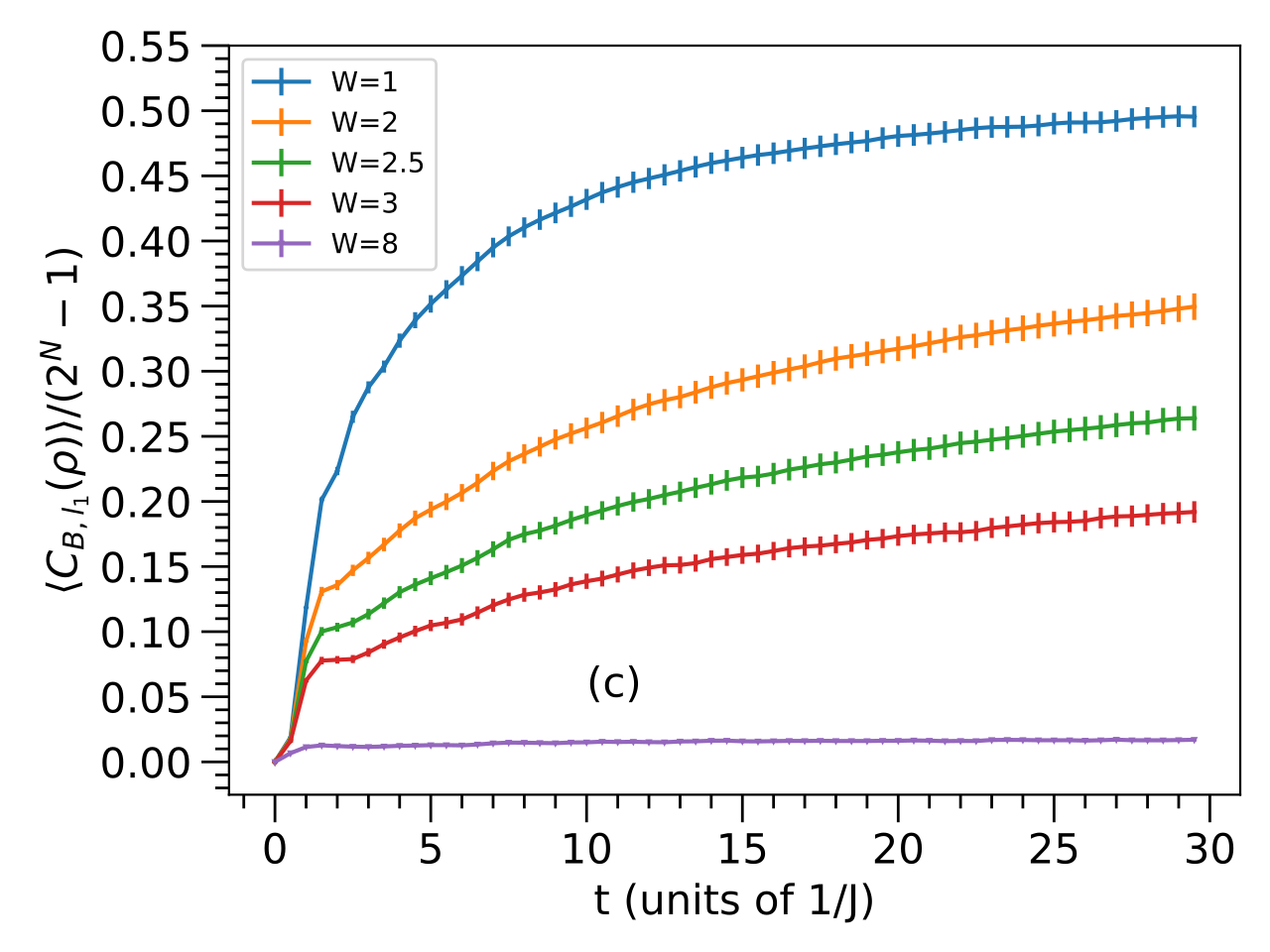}
	\caption{Time evolution of the average normalized coherence
          $\langle C_{B,l_1}(\rho)/(2^N-1) \rangle$ starting from: (a)
          the maximal coherent state $\ket{\Psi} = \frac{1}{\sqrt{d}}
          \sum_{i=1}^{d}\ket{i}$ in the computational basis, (b) the
          state $\ket{\Psi} = \ket{\uparrow\uparrow\ldots\uparrow}$,
          and (c) the state $\ket{\Psi} =
          \ket{\uparrow\downarrow\uparrow\downarrow\ldots}$. Here,
          $N=12$ and 200 disorder realizations are employed.}
	\label{fig:3}
\end{figure}

Now consider a situation away from equilibrium, e.g., a quantum
quench. After an initial preparation, we let the state evolve
unitarily under the Hamiltonian in Eq. (\ref{eq:H}) for different
strengths of the disorder $W$. In the ergodic phase, the long-time
evolution should take the state to equilibrate as a thermal ensemble
of the eigenstates of the Hamiltonian. Since these are very
delocalized in the eigenbasis of the local spins -- that is, in the
computational basis -- we expect that evolution under the ETH
Hamiltonian will have more of both coherence and decoherence power
than that of the MBL Hamiltonian. We prepare the initial state as
either (i) the maximally coherent state $\ket{\Psi} = {d}^{-1/2}
\sum_{i=1}^{d} \ket{i}$ (in the computational basis), in which case
the time evolution will decohere the state; or (ii) an incoherent
state, that is, any basis state in the computational basis. We use two
different incoherent states to make sure that the behavior of
coherence is independent of the initial energy of the system. The
results are shown in Fig. \ref{fig:3}. We see that the dynamics
induced by the ETH and MBL Hamiltonians are strikingly different in
terms of the coherence and decoherence power.  The ETH Hamiltonian
decoheres in a more efficient way a very coherent state, and, at the
same time, it is capable of building up more coherence from an
incoherent state.

Hence from studying the dynamics of quantum coherence for different
initial states, we confirm quite clearly that the MBL phase retains
the memory of the initial
state.\cite{qrevmbl,2019arXiv190810435G,ehudre}

\section{Localizable coherence}

In a quantum many-body system the Hamiltonian is the sum of local
terms, and local terms have support on local Hilbert spaces, e.g., the
spins. The total Hilbert space $\mathcal H =\otimes_i\mathcal H_i$ is
the tensor product of the local Hilbert spaces. In other words,
quantum many-body systems create a tensor product structure. Following
Ref. \onlinecite{loc_coh}, we want to quantify the coherence that is
localizable in a subsystem $S$ comprising a subset of all the
spins. For this purpose, we adopt the bipartition $\mathcal{H} =
\mathcal{H}_{S} \otimes \mathcal{H}_{R} $ ("system" and "rest") with
$\mbox{dim}(\mathcal{H}) = d = d_{S}\, d_{R}$. We then localize
coherence in the subsystem $S$ by performing a measurement on $R$. The
latter step consists of the following. Let $B_{R} :=
\{\ket{i}\}_{i=1}^{d_{R}}$ be some preferred basis in the subsystem,
where $\omega_{i} := \ket{i}\bra{i}$ form a complete set of rank-one
projectors over $\mathcal{H_{R}}$. A projective measurement on
$\mathcal{H_{R}}$ transforms a density matrix $\rho$ to a tensor
product state of the form
\begin{equation}
\rho_{i}^{\prime} = \frac{\text{Tr}_{R}\, (\rho\,
I_S\otimes\omega_{i})} {\text{Tr}\, (\rho\, I_S\otimes\omega_{i})}
  \otimes \omega_{i}.
\end{equation}
Each $\rho_{i}$ is obtained with the probability $p_{i} = \text{Tr}\,
(\rho\, I_S \otimes \omega_{i})$. One can then trace out the system
$R$ without having the state decohere and compute the coherence in $S$
in any basis of the system $B_{S}$, now described by
\begin{equation}
\rho_{S,i}^{\prime} = \text{Tr}_{R}\, \rho_{i}^\prime =
\frac{\text{Tr}_{R}\, (\rho\, \omega_{i})}{\text{Tr}(\rho\,
  \omega_{i})}.
\end{equation}
Finally, the average coherence in the post-measurement states of the
system can be defined as
\begin{equation}
C_{\rm avg}(\rho) := \sum_{i=1}^{d_{R}} p_{i}\,
C_{B_S}(\rho_{S,i}^{\prime}).
\end{equation}
The calculation of the above quantity is carried out using matrix
product states (MPS).\cite{mps1} The protocol of measurement on MPS
was first discussed by Popp and coworkers \cite{Le1} in the context of
localizable entanglement. Here we extend that formalism and calculate
the average local coherence for a particular subsystem.

We again consider the disordered Heisenberg spin ${1}/{2}$ in
Eq. (\ref{eq:H}) as a model Hamiltonian. We prepare the initial state
in an incoherent state and let it evolve. For the time-evolved state,
we calculate the localizable coherence in a subsystem consisting of
two blocks $(A,B)$ each consisting of two spins placed at a distance
$d(A,B)$ from each other. Our goal is to show that whereas the ergodic
delocalized phase should be insensitive to $d(A,B)$, in the MBL phase
the localizable coherence should be higher when the two blocks are
closer together. In order to localize coherence in the $(A,B)$ blocks,
we perform projective measurements in the rest of the system. Let us
describe the procedure for the projection in the MPS formalism. Here
we consider two blocks to be separated by three spins, $d(A, B)=3$,
but we can use similar methods for other separations. The exact
quantum state of the $N$-spin system is represented by the so-called
MPS,
\begin{equation}
\ket{\Psi} = \sum_{x_{N} = \uparrow,\downarrow} \cdots \sum_{x_{1} =
  \uparrow\downarrow} M_{N}^{x_{N}} \cdots M_{1}^{x_{1}} \ket{x_{N}
  \cdots x_{1}}.
\end{equation}
Here we will consider the localized coherence between two blocks each
consisting of two spins and separated by distance $d(A,B)=3$. Block
$A$ consists of matrices $M_{N/2-2}^{x_{N/2-2}}$ and
$M_{N/2-1}^{x_{N/2-1}}$. Block $B$ consists of matrices
$M_{N/2+2}^{x_{N/2+2}}$ and $M_{N/2+3}^{x_{N/2+3}}$. We calculate all
possible projectors on the rest of the system which is given by the
tuple $\{s\} = \{x_{N},x_{N-1},\cdots,x_{1}\} -
\{x_{N/2-2},x_{N/2-1}\}-\{x_{N/2+2},x_{N/2+3}\}$, consisting of $N-4$
spins. 
pure state after any projection can be written as
\begin{widetext}
\begin{eqnarray}
\ket{\phi_{\{s\}}} & = & \braket{\{s\}}{\Psi} \\ & = &
\sum_{x_{N/2-2}=\uparrow,\downarrow}
\sum_{x_{N/2-1}=\uparrow,\downarrow}
\sum_{x_{N/2+2}=\uparrow,\downarrow}
\sum_{x_{N/2+3}=\uparrow,\downarrow} R \cdot M_{N/2+3}^{x_{N/2+3}}\,
M_{N/2+2}^{x_{N/2+2}} \cdot Q \cdot\ M_{N/2-1}^{x_{N/2-1}}\,
M_{N/2-2}^{x_{N/2-2}} \cdot P \nonumber \\ & & \times
\ket{x_{N/2+3},x_{N/2+2}} \ket{x_{N/2-1},x_{N/2-2}},
\end{eqnarray}
\end{widetext}
where the three auxiliary matrices $R$, $Q$ and $P$ are defined as
following:
\begin{equation}
R = \sum_{\{x_{N},\ldots, x_{N/2+4}\}} M_{N}^{x_{N}} \cdot
M_{N-1}^{x_{N-1}} \cdots M_{N/2+4}^{x_{N/2+4}},
\end{equation}
\begin{equation}
Q = \sum_{\{x_{N/2 +1},x_{N/2}\}} M_{N/2+1}^{x_{N/2+1}} \cdot
M_{N/2}^{x_{N/2}}.
\end{equation}  
and
\begin{equation}
P = \sum_{\{x_{N/2 -3},\ldots,x_{1}\}} M_{N/2-3}^{x_{N/2-3}} \cdots
M_{1}^{x_{1}}.
\end{equation}
$R$,$Q$ and $P$ are computed by carrying out the matrix
multiplications for each tuple $ \{x_{N},\cdots, x_{N/2+4}\} = s_1
,\{x_{N/2 +1},x_{N/2}\} = s_2$, and $\{x_{N/2 -3},\cdots,x_{1}\} =
s_3$ respectively. There are total $2^{N-4}$ possible combinations for
$s_1$, $s_2$, and $s_3$ combined, each of which corresponds to a
different projector. The probability of a specific projector is then
given by
\begin{equation}
\text{Pr}(\{s\}) = \braket{\phi_{\{s\}}}{\phi_{\{s\}}} =
|{\braket{\{s\}}{\Psi}}|^2,
\end{equation}
and the density matrix corresponding to the projected pure state is
\begin{equation}
\rho(\{s\}) = \frac{1} {\text{Pr} (\{s\})}\ket{\phi_{\{s\}}}
\bra{\phi_{\{s\}}}.
\end{equation}
The average local coherence of the two blocks is then computed
according to the expression
\begin{equation}
\text{C}_{l_{2}}(\text{avg}) = \sum_{i=1}^{2^{N-4}}
\text{Pr}_{i}(\{s\})\times \text{C}_{l_{2}}(\rho_{i}(\{s\})).
\end{equation}
%

\begin{figure}
  \centering
  \includegraphics[width = 3in]{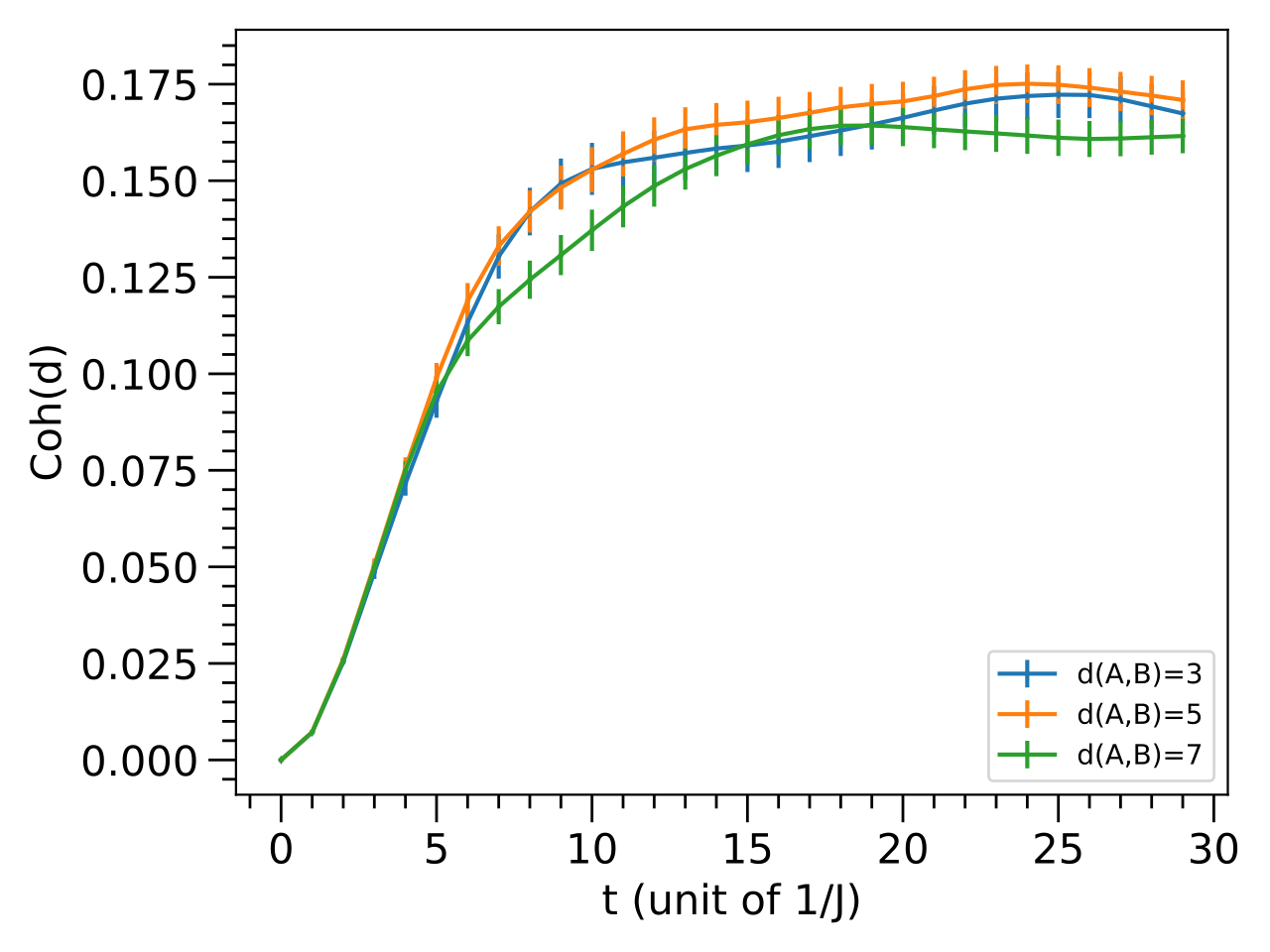}
  \includegraphics[width = 3in]{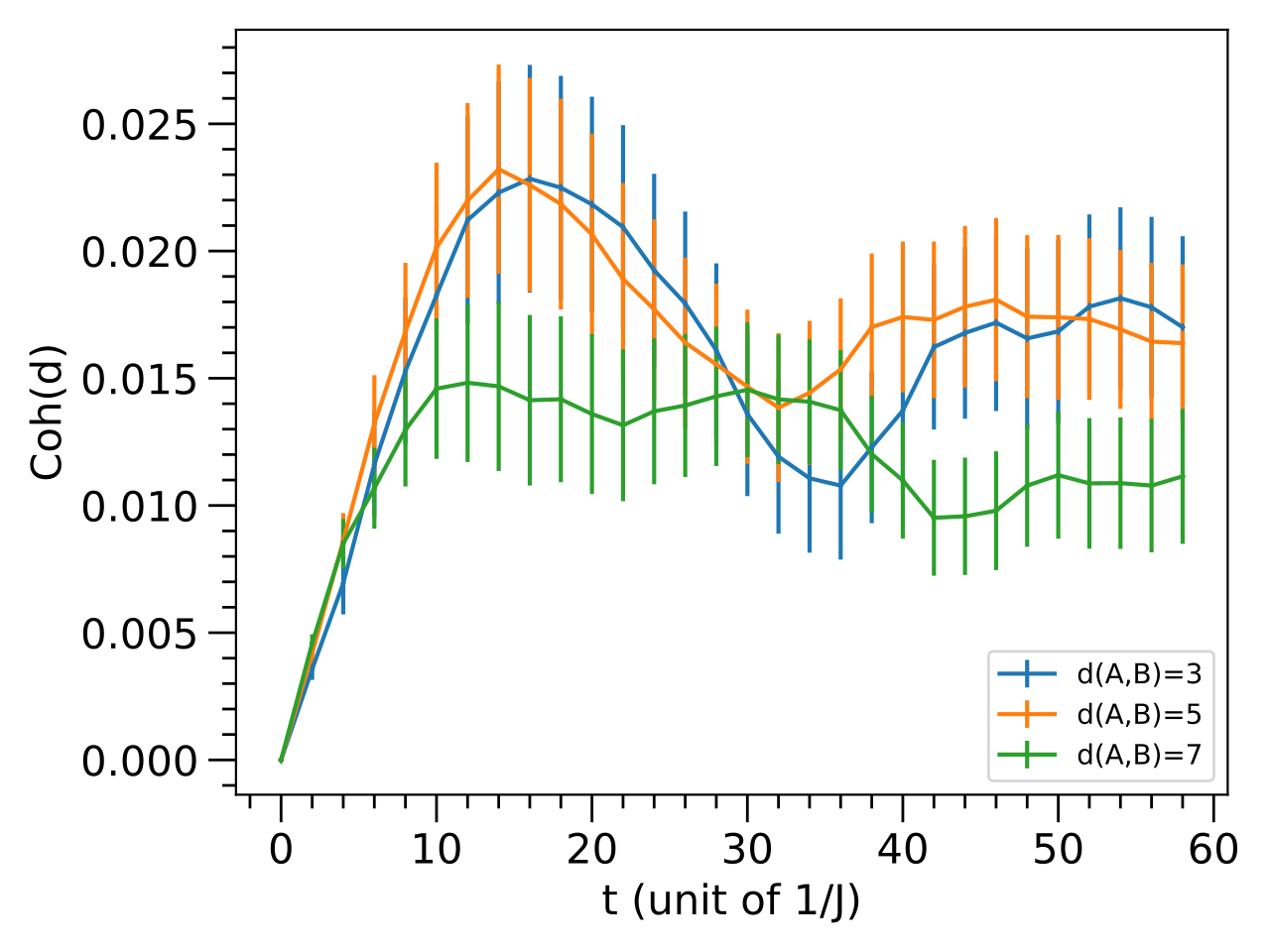}
  \caption{Average localizable coherence after a quantum quench with
    (a) the ETH Hamiltonian ($W=1$) and (b) the MBL Hamiltonian
    ($W=10$). The quantity $\langle \mbox{Coh} (d) \rangle$ is
    computed for two blocks $A$ and $B$ of two spins each at different
    distance $d(A,B)=3,5,\ldots$. The total number of spins is
    $N=14$. The initial state is the product state
    $\ket{\downarrow\downarrow...\downarrow}$. The $l_2$ norm of
    coherence is evaluated in the computational basis. The results
    represent an average over $480$ disordered samples. In case of the
    MBL Hamiltonian (b) we perform the quench for a longer time to
    specify the nature of distance dependence of average localizable
    coherence over longer timescale.}
\label{fig:4}
\end{figure}

To obtain the correlation of local coherence among these two-spin
blocks one need to subtract the effect of these individual blocks. An
effective way to do that is to calculate the local coherence of the
two-spin blocks in different locations of the disordered spin chain,
while considering the appropriate set of projective measurements on
the respective Hilbert spaces, and then take an average over the
results. One then subtract the calculated average coherence of the
individual blocks from the local coherence of the two two-spin blocks
to define $\mbox{Coh}(d)$ as localizable coherence, namely,
\begin{equation}
\text{Coh(d)} = \text{C}_{l_{2}}(\text{avg}) - \frac{1}{4}
\sum_{i=1}^4 \text{C}_{l_{2}}(p_i).
\end{equation}
Here, $\text{C}_{l_{2}}(p_i)$ refers to the coherence of the
individual two-spin blocks in several different locations along the
spin chain.


In order to compute the time evolution after the quantum quench we
utilize the time-evolving block decimation (TEBD)
method.\cite{tebd1,PAECKEL2019167998} For the TEBD, we have used a
second order Suzuki-Trotter decomposition with a time step $\delta t =
0.1$ and open boundary condition. We let the bond dimension increase
to the maximum ($D=2^{N/2}$), which in case of \ref{fig:4} is 128,
during time evolution. The time evolution reveals an important feature
of the local structure of the wave function in the ETH or MBL
phase. In ETH the many-body wave function is extended, resulting in
distance-independent behavior of the average local coherence between
different blocks, which is clearly shown in Fig. \ref{fig:4}a. In
contrast, in MBL we can see that the average local coherence between
two blocks decreases with distance when they are farther apart than
the localization length (see Fig. \ref{fig:4}b). Considering these
results, we can say that the maximum local coherence of two blocks is
higher in ETH than in the MBL phase. Since all the coherence has been
measured in the computational basis, the lower local coherence in MBL
indicates the localized structure of the wave function in the Hilbert
space.

\section{Conclusions and outlook}

In this paper, we show that measures of coherence are effective in
distinguishing the ergodic (ETH) and many-body localized (MBL) phases
and their dynamics after a quantum quench. In particular, we show that
the standard deviation of the coherence and the entropy of coherence
for a high-energy eigenstate mark the localization transition. We also
show that the time evolution of the coherence characterizes the
different dynamics of the two phases.  We then utilize a notion of
correlation of coherence based on the localizable coherence introduced
in Ref. \onlinecite{loc_coh}, to show that the ergodic phase is
insensitive to the distance between the subsystems, while it decays
for the localized phase.

We conclude that localizable coherence can be a useful instrument in
the investigation of quantum many-body systems. For example, one could
look at the fluctuations of this quantity as a probe for scrambling
and the onset of chaotic behavior in a closed quantum
system.\cite{otocANDchaos, otoc2, otocANDscr, scrINbh} Moreover, one
can think of studying in this way topological phases, as the coherence
localizable in the topological degrees of freedom should be more
robust after a quantum quench \cite{topoquench} compared to the one
localizable to local topologically trivial subsystems. Finally, as
coherence is a more experimentally accessible quantity
\cite{Yuan2020,Cohmeasure1,Cohmeasure2} compared to other quantities
used to probe into quantum many-body dynamics such as entanglement
entropy \cite{Islam2015,Lukin256}, these results should be of wide
interest to the community of quantum many-body physics.

S.D. and E.R.M. acknowledge partial financial support from NSF Grant
No. CCF-1844434.


\bibliographystyle{apsrev4-1} 

\bibliography{refs}


\end{document}